\documentclass[12pt]{article}
\usepackage{amsfonts}
\usepackage{bm}
\usepackage{amsmath}
\usepackage{epsfig}

\newcommand{\bmat}{\left(\begin{array}}
\newcommand{\emat}{\end{array}\right)}

\def\gtrsim{\mathrel{\raise.3ex\hbox{$>$\kern-.75em\lower1ex\hbox{$\sim$}}}}

\def\-{\hphantom{-}}

\def\s2{\frac{1}{\sqrt2}}

\def\mg{m_{3/2}}
\def\mg2{m^2_{3/2}}

\def\Dsl{\,\raise.15ex\hbox{/}\mkern-13.5mu D} 

\def\be{\begin{equation}}
\def\ee{\end{equation}}
\def\bea{\begin{eqnarray}}
\def\eea{\end{eqnarray}}

\newcommand{\nn}{\nonumber}

\begin{document}

\pagestyle{plain}

\makeatletter
\@addtoreset{equation}{section}
\makeatother
\renewcommand{\theequation}{\thesection.\arabic{equation}}
\pagestyle{empty}
\begin{center}
\ \

\vskip .5cm

\LARGE{\LARGE\bf T-duality and $\alpha'$-corrections \\[10mm]}
\vskip 0.3cm
\large{Diego Marqu\'es$^\dag$ and Carmen A. Nu\~nez$^{\dag\, *}$
 \\[6mm]}
{\small\it  $^\dag$ Instituto de Astronom\'ia y F\'isica del Espacio (IAFE-CONICET-UBA)\\ [.3 cm]}
{\small\it  $^*$ Departamento de F\'isica, FCEyN, Universidad de Buenos Aires (UBA) \\ [.5 cm]}

{\small \verb"{diegomarques, carmen}@iafe.uba.ar"}\\[1cm]

\small{\bf Abstract} \\[0.5cm]\end{center}

{\small We construct an $O(d,d)$ invariant universal
  formulation of the first-order $\alpha'$-corrections
  of the string effective actions involving
  the dilaton, metric and two-form fields.
   Two free
   parameters  interpolate between
   four-derivative terms that are even and odd with respect to a $Z_2$-parity
   transformation that changes the sign of the two-form field.
  The $Z_2$-symmetric model reproduces the closed
  bosonic string, and the heterotic
  string  effective action
  is obtained through a $Z_2$-parity-breaking choice of parameters.
  The theory is an extension of the generalized frame formulation of Double
  Field Theory, in which the
  gauge transformations are deformed by a first-order generalized
  Green-Schwarz transformation. This deformation defines a duality covariant gauge principle that requires and fixes the four-derivative terms. We discuss the
  $O(d,d)$ structure of the theory and the (non-)covariance of the required
  field redefinitions.
}

\newpage
\setcounter{page}{1}
\pagestyle{plain}
\renewcommand{\thefootnote}{\arabic{footnote}}
\setcounter{footnote}{0}

\tableofcontents

\section{Introduction} \label{SEC:Intro}

T-duality symmetric (double) field theories describing the
supergravity limits of string theory were originally constructed in
\cite{Siegel:1993xq}-\cite{Hohm:2010xe} and have been
studied in many recent papers (for details and references see \cite{reviews}).
Since
T-duality is a symmetry
of the string effective actions to all orders in $\alpha'$ \cite{sen},
some effort has been devoted towards
developing an $O(d,d)$ invariant formulation of the
higher order contributions. These higher derivative corrections
  are important in
 string phenomenology and
cosmology and in string theoretic studies of black hole entropy, and
such  formulation
could be useful in order to understand if/how T-duality mixes different
orders, and  could hopefully become a tool to compute or provide clues on
 the $\alpha'$-corrections.

Various methods  have been used in the early times of string theory
to construct the (super)gravity limits and their higher-derivative corrections.
The first calculations
used the scattering amplitudes of the massless particles in
the tree (or
classical) approximation of the string perturbation theory and effective Lagrangians were constructed to
reproduce this S-matrix \cite{smatrix}.
The lagrangians  are not  unique because covariant
redefinitions of
the fields  do not affect the scattering amplitudes.
Later it was realized that
the $\beta$-functions of the non-linear $\sigma$-model describing string
theory on background fields could be identified
with the equations of motion for the
massless string fields \cite{sigma}.
The
$\beta$-functions
depend on the definition of the couplings and on the renormalization
prescription. Thus the effective action whose equations of motion reproduce
them is not  unique either. Fermions cannot be easily incorporated in these approaches,
and then other
methods were developed which
take supersymmetry as the starting point \cite{ssp}.
These constructions were useful to display some symmetries of the effective
actions
that had
not been previously appreciated.

There has also been a fair amount of work to understand the duality structure of the $\alpha'$-corrections.
Time ago, K. Meissner
showed in \cite{Meissner:1996sa}
that, when dimensionally reduced to one
dimension,
the $\alpha'$-corrections in the
closed bosonic string
can be expressed
solely in terms of the
duality invariant dilaton field and the generalized metric,
which is an $O(d, d)$ group element (see also \cite{Kaloper:1997ux}). The price to pay is that the components of the generalized metric involve non-covariant derivatives of the fields.
So, while the string effective actions are defined up to
covariant field redefinitions,
it appears that non-covariant field redefinitions are necessary
in order to  make the $O(d,d)$ symmetry manifest.
In other words,
the fields that
 behave covariantly under diffeomorphisms and Lorentz transformations are not good candidates
 to become components of $O(d,d)$ multiplets. Instead,
 fields that  transform as usual (i.e. \`a la Buscher \cite{Buscher:1987qj})
 under
 T-dualities, and more generally under $O(d,d)$,
 involve non-covariant redefinitions.
 A similar result was obtained
for the heterotic string in \cite{Bergshoeff:1995cg}, where
the $O(d,d)$ friendly
fields were obtained through a Lorentz non-covariant redefinition of the metric in terms of the spin connection with
torsion (a similar result involving gauge fields was recently found in \cite{Hohm:2014sxa}). Such  redefinition had been previously considered in \cite{ht},
where it was shown that the usual Green-Schwarz mechanism of anomaly
cancellation \cite{Green:1984sg} is only consistent with worldsheet
supersymmetry if the metric is non-covariantly redefined.
The resulting Lorentz non-singlet metric then transforms similarly to the
heterotic two-form field, which is also a Lorentz non-singlet.

Recently, a method for completing higher derivative corrections
was proposed in \cite{godazgar} using
duality symmetries. It is based on the observation
that duality symmetries in the reduced theory highly constrain the form of
the unreduced theory.
This method was applied
to the closed bosonic string  and the full effective action to order
$\alpha'$ was obtained from
the Riemann squared
term. Also here it is necessary to include diffeomorphism non-covariant corrections in the duality covariant scalar matrix.

The tension between (generalized) diffeomorphism covariance and
T-duality was first discussed in \cite{Hohm:2011si}-\cite{Hohm:2014eba}
in the context of Double Field Theory (DFT). There, O. Hohm and B.
Zwiebach
showed that it is impossible to cast the square of the Riemann tensor
in terms of an $O(d,d)$-valued
generalized metric. After identifying
the  terms involved in the obstruction,
 they showed that a first order
 in $\alpha'$ non-covariant redefinition of the metric could cancel them.
 Such redefinition
  is precisely a
  background independent generalization of the one performed
  in \cite{Meissner:1996sa}.
  The authors then came to the
  conclusion that any $O(d,d)$ invariant formulation of the
  Riemann tensor squared must necessarily involve non-covariant gauge
  transformations of the
  $O(d,d)$ multiplets which induce non-covariant field redefinitions of their components. This idea is further supported by
  the absence of an $O(d,d)$ covariant generalized Riemann tensor that
  contains the usual Riemann tensor as a determined component (see \cite{Siegel:1993xq},\cite{Hohm:2010xe},\cite{Hohm:2011si},\cite{Jeon:2011cn},\cite{Coimbra:2011nw}).
  If such a generalized Riemann tensor existed, it would have to transform covariantly under
  the usual generalized Lie derivative. However, the absence
  signals the need for a correction to the gauge transformations
  (which in turn would require non-covariant field redefinitions).

The first example of an
$O(d,d)$ covariant $\alpha'$-corrected theory (including  gauge
transformations, bracket and action) was presented in \cite{Hohm:2013jaa}.
The $\alpha'$-contributions are odd under a $Z_2$-parity transformation that
changes the
sign of the two-form field, and then
this theory corresponds
neither to
the closed bosonic nor to the heterotic string. Being
odd under $Z_2$-parity, a
Riemann squared term is forbidden and, interestingly, the deformed
transformations induce a Green-Schwarz-like transformation of the two-form,
so
the first order contributions are purely governed by Chern-Simons terms
\cite{Hohm:2014eba}.
Later, in \cite{Hohm:2014xsa}, it was shown
 that this theory actually belongs
 to a two-parameter family of theories that interpolates between theories
 with even
 (DFT$^+$) and
 odd (DFT$^-$) parity corrections, where
 DFT$^+$ corresponds to the closed bosonic string
while DFT$^-$ to the theory in \cite{Hohm:2013jaa}. The gauge transformations and action were worked out to cubic order in field-perturbations, and the formulation is metric-like, so the anomalous transformation of the two-form is due to diffeomorphisms rather than Lorentz transformations.

Following a different approach, the duality structure of the $\alpha'$-corrections in the
heterotic string was
recently considered in \cite{Bedoya:2014pma}-\cite{Lee:2015kba}.
Exploiting the
symmetry
between the gauge and torsionful Lorentz connections highlighted in
\cite{Bergshoeff:1988nn},
all the first order $\alpha'$-corrections were accounted for.
The construction in \cite{Bedoya:2014pma} is based on a generalization of
the DFT formulation of the heterotic string introduced in
\cite{Hohm:2011ex}.
The gauge
and torsionful spin connections are  components of the
generalized frame, which is defined in  an extended
tangent space.
In this formulation the generalized Lie derivative
is gauged, and receives no corrections in the extended space formulation.
However, when the gauge transformations are considered
from the double space point
of view, $\alpha'$-corrections  resembling those in
\cite{Hohm:2013jaa}-\cite{Hohm:2014xsa} are induced.

In this paper we present a duality covariant gauge principle that requires and fixes the first-order contributions of a two-parameter family of theories that includes all the string effective actions. In the first part of the article we consider a two-parameter deformation of the
first order $\alpha'$-corrections in the string effective actions.
We concentrate on terms involving the metric, the Kalb-Ramond two-form
and the dilaton fields, and do not consider contributions from the gauge sector
of the heterotic string in this work.
In Section \ref{SEC:MTbR}, we compare
deformations of the four-derivative terms in the action
obtained by R. Metsaev and A.
Tseytlin from  S-matrix and $\beta-$functions calculations
in \cite{Metsaev:1987zx} with deformations
of the heterotic string effective action computed from supersymmetry
by E. Bergshoeff and M. de Roo
in
\cite{Bergshoeff:1988nn}. We prove that the deformed actions are in fact
equal up to field
redefinitions, thus generalizing the result in \cite{Chemissany:2007he}
where the agreement was shown in the case of the heterotic string.
We then construct a manifestly $O(d,d)$ invariant action which reproduces
these
four-derivative corrections. The construction presented in Section \ref{SEC:DFT}
is based on the frame-like formulation of DFT. We
introduce a first order in $\alpha'$ two-parameter deformation of the gauge transformations
of the generalized frame which takes the form of a generalized Green-Schwarz-like
transformation that induces, in particular, the anomalous transformation of the two-form field in the heterotic string.
These non-standard transformations constitute a novel duality covariant gauge principle that demands and determines the structure of the four-derivative corrections. They call
for (Lorentz) non-covariant field redefinitions,
which we discuss in detail.
Finally,  in Section \ref{Conclu}, we present the conclusions and outline future directions of research.

\section{Universal description of $\alpha'$-corrections} \label{SEC:MTbR}

The on-shell equivalence between the first order
terms in the $\alpha'$-expansion of the massless string fields effective equations
of motion  and the vanishing of
the corresponding
two-loop  terms in the Weyl anomaly coefficients of the
$\sigma$-model was verified  by R. Metsaev and A. Tseytlin in
\cite{Metsaev:1987zx}.
They showed that
the $\alpha'$-corrections involving the
metric $g_{\mu \nu}$, antisymmetric tensor $B_{\mu \nu}$ and dilaton $\phi$ fields are
parameterized by eight unambiguous coefficients which are invariant under covariant
field redefinitions and must then be determined
from  the three- and four-point scattering amplitudes of these massless states.
The results
 for the bosonic, heterotic and type II  theories
exhibit some differences. In the string frame,
four-derivative corrections are absent in the type II theories,
a Riemann squared correction plus four-derivative terms involving the two-form
field appear
in  the bosonic and heterotic theories,
and the latter contains in addition a Lorentz Chern-Simons
term in the curvature of the two-form.
While the effective action of the closed
bosonic string contains only
terms with even numbers of  Kalb-Ramond fields, and is then even under a
$Z_2$-parity transformation
that changes
the sign of $B_{\mu\nu}$,
the heterotic
string does not share this symmetry and, in particular, the
Chern-Simons
terms break the $Z_2$-parity  in the
 effective action.

The supersymmetric completion of the
$\alpha'$-corrections in the heterotic theory was obtained by
E. Bergshoeff and M. de Roo making use of a
symmetry between the gauge connection and
a spin connection with torsion \cite{Bergshoeff:1988nn}. Their results for the
bosonic sector were
shown in  \cite{Chemissany:2007he}
to coincide
with those in
\cite{Metsaev:1987zx} (modulo field redefinitions).

In this section we consider a two-parameter
deformation of the first order $\alpha'$-corrections to the string effective actions.
We first write the action in a form that makes it trivial to make contact
with the effective action  presented by R. Metsaev and A. Tseytlin in
\cite{Metsaev:1987zx}, for a specific choice of parameters.
We then rewrite it  to facilitate  comparison
with the formulation by  E. Bergshoeff and M. de Roo
in
\cite{Bergshoeff:1988nn}.
In Appendix \ref{SEC:Comparison}
we give details of the calculations allowing to go from one to the other, and
introduce the required field redefinitions and boundary terms. The two
parameters, which we denote $a$ and $b$, can be fixed to reproduce the bosonic
string $(a,b)=(-\alpha', -\alpha')$, the heterotic string
$(a,b) = (-\alpha',0)$ and (trivially)
the type II strings $(a,b)=(0,0)$ effective
actions.

\subsection{Generalized Metsaev-Tseytlin action} \label{SEC:MT}

Consider the zeroth and first-order contributions in the effective action
\be
S_{MT} = \int dx \sqrt{-g} e^{-2 \phi} \left(L^{(0)} + L^{(1)}\right) \ ,
\label{ActionMT}
\ee
where the supra-label specifies the $\alpha'$-weight.
The zeroth order (two-derivative) part of the action is just the
universal NSNS sector
\be
L^{(0)} = R - 4 \nabla_\mu \phi \nabla^\mu \phi + 4 \nabla_\mu \nabla^\mu \phi - \frac 1 {12} H^{2}  \ , \label{L0}
\ee
and the first order in $\alpha'$ (four-derivative) correction
obtained in  \cite{Metsaev:1987zx} takes the form
\bea
L^{(1)} &=& \frac {a - b} 4 H^{\mu \nu \rho}  \Omega_{\mu \nu \rho} \label{MetsaevTseytlin}\\
  && -\frac {a + b} 8 \left[ R_{\mu \nu \rho \sigma} R^{\mu \nu \rho \sigma} - \frac 1 2  H^{\mu \nu \rho}   H_{\mu \sigma \lambda} R_{\nu \rho}{}^{\sigma \lambda}+ \frac 1 {24} H^4 - \frac 1 {8} H^2_{\mu \nu} H^{2 \mu \nu} \right] \ . \nn
\eea
We use the standard notation for the components and
their definitions can be found in Appendix \ref{SEC:Conventions}.
The Metsaev-Tseytlin action
is  recovered with the following choice of parameters
\be
 \frac {a + b} 8 = - \lambda_0 \alpha' = \left\{\begin{matrix} - \frac 1 4 \alpha' \ \ \ \ {\rm bosonic \ string} \ \ \\ - \frac 1 8 \alpha'  \ \ \ \ {\rm heterotic \ string}\\ 0  \ \ \ \ \ \ \ {\rm type \ II}\ \  \ \ \ \ \ \ \end{matrix} \right. \ , \ \ \ \ \ \frac {a - b} 8 = \left\{\begin{matrix} \ \ 0 \ \ \ \ \ \ {\rm bosonic \ string} \ \ \\ - \frac 1 8 \alpha'  \ \ \  {\rm heterotic \ string}\\ \ \ 0  \ \ \ \ \ \ {\rm type \ II} \ \ \ \  \ \ \ \ \ \ \end{matrix} \right. \ .
\ee

Notice that for the bosonic string the first term in (\ref{MetsaevTseytlin}) is absent, and only terms that contain even powers of the three-form $H$
are non-vanishing. As a result the action is symmetric under a $Z_2$-parity transformation that exchanges the sign of the Kalb-Ramond two-form
\be
Z_2(B) = - B \ ,
\ee
i.e. $Z_2(L^{(1)}) = L^{(1)}$. The heterotic string is not symmetric under this parity transformation, because in this case the first term in (\ref{MetsaevTseytlin}) changes sign. There is another interesting case, corresponding to
the choice $a + b = 0$, in which the first-order corrections are purely given
by the first term in (\ref{MetsaevTseytlin}) and are then
odd under $Z_2$-parity, i.e. $Z_2(L^{(1)}) = - L^{(1)}$. This case is very likely
related to one recently introduced in \cite{Hohm:2013jaa} and
further discussed in \cite{Hohm:2014eba}.

The action (\ref{ActionMT}) is invariant under diffeomorphisms and gauge transformations of the two-form. However, Lorentz invariance requires the non-standard Lorentz transformation of the two-form
\be
\delta_\Lambda B^{\rm MT}_{\mu \nu} = - \frac 1 2 (a - b) \partial_{[\mu} \Lambda_a{}^b \omega_{\nu]b}{}^a \ , \label{varBMT}
\ee
which is necessary for anomaly cancellations in the Green-Schwarz mechanism. Clearly, this transformation is not present in the bosonic string, but appears as expected in the heterotic string.

\subsection{Generalized Bergshoeff-de Roo action} \label{SEC:BR}

Consider now the following  action
\bea
S_{BR} &=& \int dx \sqrt{- g} e^{- 2 \phi} \left(R - 4 \nabla_\mu \phi \nabla^\mu \phi + 4 \nabla_\mu \nabla^\mu \phi - \frac 1 {12} \widetilde H^{\mu \nu \rho} \widetilde H_{\mu \nu \rho} \right. \nn\\
&& \left. \ \ \ \ \ \ \ \ \ \ \ \ \ \ \ \ \ \ \ \ \ \ + \frac a 8 R^{(-)}_{\mu \nu a}{}^b R^{(-)\mu \nu}{}_b{}^a + \frac b 8 R^{(+)}_{\mu \nu a}{}^b R^{(+)\mu \nu}{}_b{}^a \right) \ , \label{BergshoeffdeRoo}
\eea
where
\be
\widetilde H_{\mu \nu \rho} = H_{\mu \nu \rho} - \frac 3 2 a \Omega^{(-)}_{\mu \nu \rho} + \frac 3 2 b \Omega^{(+)}_{\mu \nu \rho} \ . \label{tH}
\ee
The case $(a,b) = (- \alpha',0)$ corresponds to the heterotic string, and coincides with the bosonic sector of the effective action as presented in \cite{Bergshoeff:1988nn}. For this choice of parameters, this action was shown in \cite{Chemissany:2007he} to coincide (modulo field redefinitions and boundary terms) with the Metsaev-Tseytlin action given above in (\ref{ActionMT}) with the same choice of parameters. In Appendix \ref{SEC:Comparison} we generalize the identification, making it valid for any choice of parameters. The field redefinitions involved in the computations are mostly diffeomorphism and Lorentz covariant, except for a Lorentz non-covariant redefinition of the two-form field given by (see (\ref{deltaBnoncovMTvsBR}))
\be
B^{MT} = B^{BR} + \Delta B \ , \ \ \ \ \ \ \ \Delta B_{\mu \nu} = - \frac 1 4 (a + b) H_{[\mu}{}^{a b} \omega_{\nu] a b} \ . \label{NonCovFRB}
\ee

The action (\ref{BergshoeffdeRoo}) is invariant under diffeomorphisms and gauge transformations of the two-form. However,
Lorentz invariance again requires a non-standard Lorentz transformation of the two-form
\bea
\delta_\Lambda B^{\rm BR}_{\mu \nu} &=& - \frac a 2  \partial_{[\mu} \Lambda_a{}^b \omega^{(-)}_{\nu]b}{}^a + \frac b 2  \partial_{[\mu} \Lambda_a{}^b \omega^{(+)}_{\nu]b}{}^a \nn \\
&=& - \frac  1 2 (a - b) \partial_{[\mu} \Lambda_a{}^b \omega_{\nu]b}{}^a + \frac 1 4 (a + b) \partial_{[\mu} \Lambda_a{}^b H_{\nu]b}{}^a \ , \label{varBBR}
\eea
necessary for anomaly cancelations in the Green-Schwarz mechanism. Notice that the field redefinition (\ref{NonCovFRB}) eliminates the last term
in the transformation (\ref{varBBR}) of the two-form, making it equal to
that in (\ref{varBMT}).

\section{$\alpha'$-corrections in Double Field Theory} \label{SEC:DFT}

In this section we introduce the $O(d,d)$ invariant frame-like formulation of Double Field Theory (DFT) that reproduces the two-parameter
deformed action introduced above.
The zeroth order frame-like theory was introduced in \cite{Siegel:1993xq}, further explored in \cite{Hohm:2010xe}, and here we will mostly follow the conventions of \cite{Geissbuhler:2013uka}. Our original contribution here is a two-parameter first-order in $\alpha'$ deformation of the gauge transformations of the generalized frame, that takes the form of a generalized Green-Schwarz-like transformation that induces in particular the anomalous Lorentz transformation of the two-form. We  first introduce the fields, their transformation properties and closure of the algebra, and we finally write an invariant action to first order in $\alpha'$. Then, we  show that the action exactly reproduces the two-parameter  action
(\ref{BergshoeffdeRoo}), when taking the standard solution of the strong constraint together with a compatible parameterization of fields.

\subsection{Generalized fields, projectors and fluxes}

The DFT action is invariant under global $G = O(d,d)$ transformations, local ``double-Lorentz'' $H=O(1,d-1) \times O(d-1,1)$ transformations, and infinitesimal
generalized diffeomorphisms generated by a generalized Lie derivative $\widehat {\cal L}$. A constant symmetric and invertible $G$-invariant metric $\eta_{M N}$
raises and lowers the indices that are rotated by $G$ (which we label $M,N,\dots$). In addition, there are two constant symmetric and invertible $H$-invariant
metrics $\eta_{A B}$ and ${\cal H}_{A B}$. The former is used to raise
and lower the indices that are rotated by $H$ (which we label $A,B,\dots$), and the
latter is
constrained to satisfy
\be
{\cal H}_{A C} \eta^{C D} {\cal H}_{D B} = \eta_{A B} \ . \label{flatconstraint}
\ee
The three metrics are invariant under the action of $\widehat{\cal L}$, $G$ and $H$.

The theory is defined on a double space, in which derivatives $\partial_M$ belong to the fundamental representation of $G$.
However, a strong constraint
\be
\partial_M \partial^M \dots = 0 \ , \ \ \ \ \ \partial_M \dots \ \partial^M \dots = 0 \ , \label{StrongConstraint}
\ee
restricts  the fields and gauge parameters,
the dots representing arbitrary products of them. While the generalized Lie derivative is generated by an infinitesimal generalized parameter $\xi^M$ that takes values in the fundamental representation of $G$, $H$-transformations are generated by an infinitesimal parameter $\Lambda_{A}{}^B$. The latter is constrained by the fact that $\eta_{A B}$ and ${\cal H}_{A B}$ must be $H$-invariant
\be
\delta_\Lambda \eta_{A B} = \eta_{C B} \Lambda^C{}_A + \eta_{A C} \Lambda^C{}_B = 0 \ , \ \ \ \ \ \delta_\Lambda {\cal H}_{A B} = {\cal H}_{C B} \Lambda^C{}_A + {\cal H}_{A C} \Lambda^C{}_B = 0 \ . \label{constraintsLambda}
\ee

The fields of the theory are a generalized frame $E_M{}^A$ and a generalized dilaton $d$. The generalized frame relates the metric $\eta_{A B}$ with $\eta_{M N}$, and the metric ${\cal H}_{A B}$ with the so-called generalized metric ${\cal H}_{M N}$
 \be
\eta_{M N} = E_M{}^A \eta_{A B} E_N{}^B     \ , \ \ \ \ \ \      {\cal H}_{M N} = E_M{}^A {\cal H}_{A B} E_N{}^B \ . \label{flattocurve}
\ee
As a result of (\ref{flatconstraint}), the generalized metric is constrained to be $G$-valued
\be
{\cal H}_{M P} \eta^{P Q} {\cal H}_{Q N} = \eta_{M N} \ . \label{curveconstraint}
\ee
It is important to point out that the generalized fields and gauge parameters are allowed to receive
corrections that respect the constraints. We will give concrete expressions for the first order corrections to their components later.

Since the generalized metric is constrained by (\ref{curveconstraint}), one can define the following projectors
\be
P_{M N} = \frac 1 2 \left( \eta_{M N} - {\cal H}_{M N}\right) \ , \ \ \ \ \ \bar P_{M N} = \frac 1 2 \left( \eta_{M N} + {\cal H}_{M N}\right) \ , \label{projectors}
\ee
which satisfy the following identities
\be
P_M{}^Q P_Q{}^N = P^N_M \ , \ \ \ \ \ \bar P_M{}^Q \bar P_Q{}^N = \bar P^N_M \ , \ \ \ \ \ P_M{}^Q \bar P_Q{}^N = 0 \ .
\ee
In complete analogy, one can define these projectors in flat indices
\be
P_{A B} = \frac 1 2 \left( \eta_{A B} - {\cal H}_{A B}\right) \ , \ \ \ \ \ \bar P_{A B} = \frac 1 2 \left( \eta_{A B} + {\cal H}_{A B}\right) \ , \label{projectors}
\ee
which satisfy analogous identities
\be
P_A{}^C P_C{}^B = P^B_A \ , \ \ \ \ \ \bar P_A{}^C \bar P_C{}^B = \bar P^B_A \ , \ \ \ \ \ P_A{}^C \bar P_C{}^B = 0 \ .
\ee
Another useful identity is
\be
P_{M}{}^N E_N{}^A = E_M{}^B P_B{}^A \ , \ \ \ \ \ \ \bar P_{M}{}^N E_N{}^A = E_M{}^B \bar P_B{}^A \ .
\ee
We will use the barred-index notation to denote projections
\be
P_M{}^N V_N = V_{\underline{M}} \ , \ \ \ \ \ \bar P_M{}^N V_N = V_{\overline{M}} \ ,
\ee
and  the following convention for (anti-)symmetrization of
barred-indices
\be
V_{(\underline{M}} W_{\overline{N})} = \frac 1 2 \left( V_{\underline{M}} W_{\overline{N}} + V_{\underline{N}} W_{\overline{M}}\right) \ , \ \ \ \ \ V_{[\underline{M}} W_{\overline{N}]} = \frac 1 2 \left( V_{\underline{M}} W_{\overline{N}} - V_{\underline{N}} W_{\overline{M}}\right)\ ,
\ee
i.e.,  only the indices are exchanged and not the bars.

Important objects in the frame-like or flux-formulation of DFT are the generalized fluxes
\be
{\cal F}_{A B C} = 3 E_{M [A} \partial^M E^N{}_B E^P{}_{C]} \eta_{N P} \ ,
\ee
and
the following projections take a predominant role in the $\alpha'$-deformed theory that we will introduce
\bea
{\cal F}^{(-)}_{M A B} &=& {\cal F}_{\overline M \underline A \underline B} = \bar P_M{}^N E_N{}^C {\cal F}_{C D E} P_A{}^D P_B{}^E\, ,\\
{\cal F}^{(+)}_{M A B} &=& {\cal F}_{\underline M \overline A \overline B} = P_M{}^N E_N{}^C {\cal F}_{C D E} \bar P_A{}^D \bar P_B{}^E \ .
\eea

Let us finally discuss $Z_2$ transformations. They are generated by  matrices $Z_M{}^N$ and $Z_A{}^B$ that transform the metrics as follows
\bea
Z_2 \left(\eta_{A B}\right) &=& Z_A{}^C \eta_{C D} Z_B{}^D = - \eta_{A B} \, ,\\
Z_2 \left({\cal H}_{A B}\right) &=& Z_A{}^C {\cal H}_{C D} Z_B{}^D = {\cal H}_{A B}
\, ,\\
Z_2 \left(\eta_{M N}\right) &=& Z_M{}^P \eta_{P Q} Z_N{}^Q = - \eta_{M N} \ .
\eea
Since indices are raised and lowered with the odd $Z_2$
metrics $\eta_{M N}$ and $\eta_{A B}$, the position of the indices is essential to determine the way in which an object transforms under $Z_2$-parity. There is a canonical position of indices that renders the following objects even under
$Z_2$: $\partial_M$, ${\cal H}_{M N}$, $E_M{}^A$, ${\cal F}_{A B}{}^C$, $\xi^M$ and $\Lambda_A{}^B$. This in turn implies that the projectors are exchanged under $Z_2$, namely $Z_2(P_\bullet{}^\bullet) = \bar P_\bullet{}^\bullet$ and $Z_2(\bar P_\bullet{}^\bullet) = P_\bullet{}^\bullet$, and then
\be
Z_2\left({\cal F}^{(\pm)}_{M A}{}^B\right) = {\cal F}^{(\mp)}_{M A}{}^B \ .
\ee

\subsection{Generalized Green-Schwarz transformations}

The generalized dilaton and frame transform under generalized diffeomorphisms and $H$-transformations as
\bea
\delta d &=& \xi^P \partial_P d - \frac 1 2 \partial_P \xi^P  \ \ \ \ \Leftrightarrow \ \ \ \ \delta e^{-2d} = \partial_P \left(\xi^P e^{-2 d}\right) \, , \label{vard}\\
\delta E_M{}^A &=& \widehat {\cal L}_\xi E_M{}^A + \delta_\Lambda E_M{}^A  + \widetilde \delta_\Lambda E_M{}^A \ ,
\eea
where
the generalized Lie derivative governing infinitesimal generalized diffeomorphisms
is given by
\be
\widehat {\cal L}_\xi E_M{}^A = \xi^P \partial_P E_M{}^A + \left(\partial_M \xi^P - \partial^P \xi_M\right) E_P{}^A \ ,
\ee
and $H$-transformations split in the usual
\be
\delta_\Lambda E_M{}^A = E_M{}^B \Lambda_B{}^A \ , \label{DeltaLambdaE}
\ee
plus a novel two-parameter first-order correction
\be
\widetilde \delta_\Lambda E_M{}^A =  \left( a\, \partial_{[\underline M} \Lambda_{C}{}^{B} \, {\cal F}^{(-)}_{\overline N] B}{}^{C} -  b\, \partial_{[\overline M} \Lambda_{C}{}^{B} \, {\cal F}^{(+)}_{\underline N] B}{}^{C}\right) E^{N A} \ .\label{genGSgenframe}
\ee
The parameters $(a,\, b)$ are both of ${\cal O}(\alpha')$. This first-order correction suggests that the component fields  parameterizing the generalized fields cannot be the standard ones that transform covariantly under diffeomorphisms and Lorentz transformations. Instead, they should
correspond to first order non-covariantly redefined fields, and then the generalized fields must be $\alpha'$-corrected $E = E^{(0)} + E^{(1)}$. The same holds
 for the gauge parameter $\Lambda = \Lambda^{(0)} + \Lambda^{(1)}$. Since (\ref{genGSgenframe}) is already of ${\cal O}(\alpha')$ through $(a,b)$, only $E^{(0)}$ and $\Lambda^{(0)}$ are relevant in this part of the transformations.

For the generalized metric these transformations imply
\be
\delta {\cal H}_{M N} = \widehat {\cal L}_\xi {\cal H}_{M N} + \widetilde \delta_\Lambda {\cal H}_{M N} \ , \label{vargenmet}
\ee
with
\be
\widehat {\cal L}_\xi {\cal H}_{M N} = \xi^P \partial_P {\cal H}_{M N} + \left(\partial_M \xi^P - \partial^P \xi_M\right) {\cal H}_{P N} + \left(\partial_N \xi^P - \partial^P \xi_N\right) {\cal H}_{M P} \ ,
\ee
and
\be
\widetilde \delta_\Lambda {\cal H}_{M N} = 2 a\, \partial_{(\underline M} \Lambda_{A}{}^{B} \, {\cal F}^{(-)}_{\overline N) B}{}^{A} + 2 b\, \partial_{(\overline M} \Lambda_{A}{}^{B} \, {\cal F}^{(+)}_{\underline N) B}{}^{A} \ . \label{genGSgenmetric}
\ee
Notice that the first-order double-Lorentz transformations $\widetilde \delta_\Lambda$ in (\ref{genGSgenframe}) and (\ref{genGSgenmetric}) take the form of a generalized Green-Schwarz transformation for the generalized fields, i.e. they are structurally similar to (\ref{varBBR}). We will show in the following sections that these transformations indeed induce the Green-Schwarz transformation (\ref{varBBR}) of the two-form when the strong constraint is properly solved, plus an anomalous Lorentz transformation of the metric field, which can however be eliminated through a Lorentz non-covariant field redefinition.
Again, $\widetilde \delta_\Lambda {\cal H}$ is ${\cal O}(\alpha')$, and then also the generalized metric is $\alpha'$-corrected ${\cal H} = {\cal H}^{(0)} + {\cal H}^{(1)}$.

Regarding the transformation of the fluxes, to lowest order in $\alpha'$ they transform as
\be
\delta {\cal F}_{A B C} =  \xi^P \partial_P {\cal F}_{A B C} - 3 \left( \partial_{[A}\Lambda_{B C]} + \Lambda_{[A}{}^D {\cal F}_{B C] D}\right) \ ,
\ee
which implies that the projected generalized fluxes transform as connections to lowest order
\bea
\delta {\cal F}^{(-)}_{M A}{}^B &=& \widehat {\cal L}_\xi  {\cal F}^{(-)}_{M A}{}^B - \partial_{\overline{M}} \Lambda_{\underline{A}}{}^{\underline{B}} + {\cal F}^{(-)}_{M A}{}^C \Lambda_C{}^B - \Lambda_A{}^C {\cal F}^{(-)}_{M C}{}^B \, ,\nn \\
\delta {\cal F}^{(+)}_{M A}{}^B &=& \widehat {\cal L}_\xi  {\cal F}^{(+)}_{M A}{}^B - \partial_{\underline{M}} \Lambda_{\overline{A}}{}^{\overline{B}} + {\cal F}^{(+)}_{M A}{}^C \Lambda_C{}^B - \Lambda_A{}^C {\cal F}^{(+)}_{M C}{}^B \ , \label{varprojectedF}
\eea
with
\be
\widehat {\cal L}_\xi {\cal F}^{(\pm)}_{M A}{}^B = \xi^P \partial_P {\cal F}^{(\pm)}_{M A}{}^B + \left(\partial_M \xi^P - \partial^P \xi_M \right) {\cal F}^{(\pm)}_{P A}{}^B \ .
\ee
The fields ${\cal F}^{(\pm)}_{M A B}$ appear
in the action (to be introduced in the next section)
only in terms that are weighted with $a$ and $b$. Then, in order to prove the gauge invariance of the action to ${\cal O}(\alpha')$, only their lowest order transformations are required.

The above transformations preserve the constraints of the generalized fields (\ref{flattocurve}) and (\ref{curveconstraint}), and also close to first order
\be
\left[ \delta_{(\xi_1 \, , \, \Lambda_1)} , \delta_{(\xi_2 \, , \, \Lambda_2)} \right] = \delta_{(\xi_{21} \, , \, \Lambda_{21})} \ ,
\ee
where the ``brackets'' are given by
\bea
\xi_{12}^M \!&=&\! \left[\xi_1 \, , \, \xi_2 \right]^M_{(C)} - \frac a 2 \Lambda_{[1\, \underline A}{}^{\underline B} \partial^M \Lambda_{2]\, \underline B}{}^{\underline A} + \frac b 2 \Lambda_{[1\, \overline A}{}^{\overline B} \partial^M \Lambda_{2]\, \overline B}{}^{\overline A} \, , \label{alphaprimebracket}\\
\Lambda_{12\, A B} \!&=&\! 2 \xi_{[1}^P \partial_P \Lambda_{2] \, A B} - 2 \Lambda_{[1\, A}{}^C \Lambda_{2]\, C B} \\
\!&&\! +\, a \, \partial_{[\overline A}\Lambda_1^{\underline C \underline D} \partial_{\overline B]} \Lambda_{2 \underline D \underline C} +\, a \, \partial_{[\underline A}\Lambda_1^{\underline C \underline D} \partial_{\underline B]} \Lambda_{2 \underline D \underline C} - \, b \, \partial_{[\underline A}\Lambda_1^{\overline C \overline D} \partial_{\underline B]} \Lambda_{2 \overline D \overline C} - \, b \, \partial_{[\overline A}\Lambda_1^{\overline C \overline D} \partial_{\overline B]} \Lambda_{2 \overline D \overline C}\ ,\nn
\eea
and the C-bracket is defined as
\be
\left[\xi_1 \, , \, \xi_2 \right]^M_{(C)} = \xi_1^P \partial_P \xi_2^M - \xi_2^P \partial_P \xi_1^M - \frac 1 2 \xi_1^P \partial^M \xi_{2P} + \frac 1 2 \xi_2^P \partial^M \xi_{1P} \ .
\ee

It is interesting to note that due to the constraints (\ref{constraintsLambda}), the $\alpha'$-corrected bracket (\ref{alphaprimebracket}) can be re-written as
\be
\xi_{12}^M = \left[\xi_1 \, , \, \xi_2 \right]^M_{(C)} + \frac 1 2 \left(\gamma^{(+)} {\cal H}^{A B} - \gamma^{(-)} \eta^{A B}\right) \eta^{C D} \Lambda_{[1 A C} \partial^M \Lambda_{2]B D} \ ,
\ee
where
\be
\gamma^{(\pm)} = - \frac {a \pm b} 2 \ .
\ee
This re-writing allows to facilitate comparison with the deformed brackets introduced in \cite{Hohm:2014xsa}. There, the parameters $\gamma^{(-)}$ and $\gamma^{(+)}$ interpolate between the odd $Z_2$-parity theory $DFT^{-}$ in \cite{Hohm:2013jaa} obtained through the choice $(\gamma^{(+)}, \gamma^{(-)}) = (0,1)$ when $\alpha' = 1$, and the even $Z_2$-parity theory $DFT^+$ obtained through the choice $(\gamma^{(+)}, \gamma^{(-)}) = (1,0)$ corresponding to the closed bosonic string. It is not evident a priori that  both approaches can be compared because here the deformations are due to double Lorentz parameters $\Lambda_{A B}$, and in \cite{Hohm:2014xsa} are due to generalized diffeomorphisms through $K_{M N} = \partial_{M}\xi_{N} - \partial_{N}\xi_{M}$. It would be interesting to explore the relation between both approaches.

Notice that the $Z_2$-transformation of the generalized Green-Schwarz transformation (\ref{genGSgenmetric}) is
\be
Z_2 \left(\widetilde \delta_\Lambda {\cal H}_{M N}\right) = 2 b\, \partial_{(\underline M} \Lambda_{A}{}^{B} \, {\cal F}^{(-)}_{\overline N) B}{}^{A} + 2 a\, \partial_{(\overline M} \Lambda_{A}{}^{B} \, {\cal F}^{(+)}_{\underline N) B}{}^{A} \ ,
\ee
so the $Z_2$-transformation effectively exchanges the parameters $a \leftrightarrow b$. Then, the transformation is even under $Z_2$-parity
when $a = b$ (which in turn implies $\gamma^{(-)} = 0$) and odd when $a = - b$ (which in turn implies $\gamma^{(+)} = 0$). Any other choice of parameters breaks
$Z_2$-parity.

\subsection{Gauge invariant action}
We now have all the ingredients to write down a gauge-invariant action to first order in $\alpha'$
\be
S = \int dX e^{-2 d}\left({\cal R} + a \, {\cal R}^{(-)} + b \, {\cal R}^{(+)} \right) \ , \label{ActionAlphaPrime}
\ee
where $\cal R$ is of course defined in the same way as the zeroth order DFT action \cite{Hull:2009mi}
\bea
{\cal R} &=& 4 {\cal H}^{M N} \partial_{M N}{d} - \partial_{M N}{{\cal H}^{M N}} - 4 {\cal H}^{M N} \partial_{M}{ d } \partial_{N}{ d } + 4 \partial_{M}{{\cal H}^{M N} } \partial_{N}{ d } \nn \\
&& + \frac 1 8  {\cal H}^{M N} \partial_{M}{ {\cal H}^{K L} } \partial_{N}{ {\cal H}_{K L} } - \frac 1 2 {\cal H}^{M N} \partial_{M}{ {\cal H}^{K L} } \partial_{K}{ {\cal H}_{N L} } \ . \label{DFTaction}
\eea
As explained, the generalized metric is $\alpha'$-corrected ${\cal H} = {\cal H}^{(0)} + {\cal H}^{(1)}$, so even if this looks like a two-derivative contribution, $\cal R$ involves four-derivative terms through the corrections to the fields. Of course, in the limit $\alpha' \to 0$ we should recover the usual un-corrected action, so $\cal R$ is a good starting point to build the ${\cal O}(\alpha')$ action. While $\cal R$ is a scalar under generalized diffeomorphisms, it fails to be
gauge invariant
under generalized Green-Schwarz transformations (\ref{genGSgenmetric}).
Then, additional contributions to the Lagrangian must be considered to compensate for this failure, which
must be scalars themselves under generalized diffeomorphisms
as well. It is in this sense that the generalized Green-Schwarz transformations constitute a gauge principle that requires and fixes the form of the $\alpha'$-corrections.
Since (\ref{genGSgenmetric}) induces terms that involve the projected generalized fluxes ${\cal F}^{(\pm)}_{M A B}$, so must the corrections to the action. In fact, one can show that the required additional first-order corrections from the projected fluxes ${\cal F}^{(-)}_{M A B}$ are given by\footnote{The full action is frame-like since it depends on the generalized frame through the generalized metric and the projected fluxes. In would be interesting to see if this hybrid formulation can be written purely in terms of generalized fluxes as in \cite{Geissbuhler:2013uka}.}
\bea
{\cal R}^{(-)} &=& - 4  {\cal F}^{(-)}_{M A B} {\cal F}^{(-) B A}_{N}  \partial^{M N}{d}
              +   \partial^{M N}\left({\cal F}^{(-)}_{M A B} {\cal F}^{(-) B A}_{N}\right)  \nn \\
              && + 4   {\cal F}^{(-)}_{M A B} {\cal F}^{(-) B A}_{N}  \partial^M{d} \partial^N{d}
              - 4   \partial^M \left({\cal F}^{(-)}_{M A B} {\cal F}^{(-) B A}_{N}\right)  \partial^N{d} \nn \\
             && - \frac 1 8   {\cal F}^{(-)}_{M A B} {\cal F}^{(-) B A}_{N} \partial^{M}{{\cal H}^{R S}} \partial^{N}{{\cal H}_{R S}}
               + \frac 1 2  {\cal F}^{(-)}_{M A B} {\cal F}^{(-)N B A} \partial^{M}{\cal H}^{R S} \partial_{R}{\cal H}_{N S} \nn \\
              && - \frac 1 4 {\cal H}^{M N} \partial_{M}{{\cal H}^{P Q}} \partial_{N}\left({\cal F}^{(-)}_{P A B} {\cal F}^{(-) B A}_{Q}\right)
              + \frac 1 2  {\cal H}^{R S} \partial_{R}\left({\cal F}^{(-)}_{M A B} {\cal F}^{(-)N B A}\right) \partial^M{{\cal H}_{S N}} \nn \\
              && + \frac 1 2  {\cal H}^{R S} \partial_{R}{{\cal H}^{P Q}} \partial_{P}\left({\cal F}^{(-)}_{S A B} {\cal F}^{(-) B A}_{Q}\right)
              + \frac 1 2 {\cal H}^{M N} \partial_{M}{{\cal F}^{(-)}_{R A B}} \partial_{N}{{\cal F}^{(-)R B A}} \nn \\
              &&  -  {\cal F}^{(-)}_{M A B} \partial^M{{\cal H}^{K L}} \partial_{K}{{\cal F}^{(-) B A}_{L}}
               -  {\cal H}^{M N} \partial_{M}{{\cal F}^{(-)}_{R A B}} \partial^{R}{{\cal F}^{(-) B A}_{N}} \nn \\
              && - 4   {\cal F}^{(-)}_{M A B} {\cal F}^{(-)N B C} \partial^M{{\cal F}^{(-)}_{N C}{}^{A}}
              +   {\cal F}^{(-)}_{M A B} {\cal F}^{(-) M}{}_{C D} {\cal F}^{(-) A C}_{P} {\cal F}^{(-) P B D} \nn \\
              && -   {\cal F}^{(-)}_{M A B} {\cal F}^{(-) M A}{}_{D} {\cal F}^{(-)}_{P E}{}^B {\cal F}^{(-) P E D} \ , \label{Rm}
\eea
and the other first-order corrections from the projected fluxes ${\cal F}^{(+)}_{M A B}$ are given by
\bea
{\cal R}^{(+)} &=& - 4  {\cal F}^{(+)}_{M A B} {\cal F}^{(+) B A}_{N}  \partial^{M N}{d}
              +  \partial^{M N}\left({\cal F}^{(+)}_{M A B} {\cal F}^{(+) B A}_{N}\right)  \nn \\
              && + 4   {\cal F}^{(+)}_{M A B} {\cal F}^{(+) B A}_{N}  \partial^M{d} \partial^N{d}
              - 4  \partial^M \left({\cal F}^{(+)}_{M A B} {\cal F}^{(+) B A}_{N}\right)  \partial^N{d} \nn \\
             && - \frac 1 8   {\cal F}^{(+)}_{M A B} {\cal F}^{(+) B A}_{N} \partial^{M}{{\cal H}^{R S}} \partial^{N}{{\cal H}_{R S}}
              + \frac 1 2  {\cal F}^{(+)}_{M A B} {\cal F}^{(+)N B A} \partial^{M}{\cal H}^{R S} \partial_{R}{\cal H}_{N S} \nn \\
              && - \frac 1 4 {\cal H}^{M N} \partial_{M}{{\cal H}^{P Q}} \partial_{N}\left({\cal F}^{(+)}_{P A B} {\cal F}^{(+) B A}_{Q}\right)
              + \frac 1 2  {\cal H}^{R S} \partial_{R}\left({\cal F}^{(+)}_{M A B} {\cal F}^{(+)N B A}\right) \partial^M{{\cal H}_{S N}} \nn \\
              && + \frac 1 2  {\cal H}^{R S} \partial_{R}{{\cal H}^{P Q}} \partial_{P}\left({\cal F}^{(+)}_{S A B} {\cal F}^{(+) B A}_{Q}\right)
              - \frac 1 2 {\cal H}^{M N} \partial_{M}{{\cal F}^{(+)}_{R A B}} \partial_{N}{{\cal F}^{(+)R B A}} \nn \\
              && +  {\cal F}^{(+)}_{M A B} \partial^M{{\cal H}^{K L}} \partial_{K}{{\cal F}^{(+) B A}_{L}}
               +   {\cal H}^{M N} \partial_{M}{{\cal F}^{(+)}_{R A B}} \partial^{R}{{\cal F}^{(+) B A}_{N}} \nn \\
              && - 4   {\cal F}^{(+)}_{M A B} {\cal F}^{(+)N B C} \partial^M{{\cal F}^{(+)}_{N C}{}^{A}}
              +  {\cal F}^{(+)}_{M A B} {\cal F}^{(+) M}{}_{C D} {\cal F}^{(+) A C}_{P} {\cal F}^{(+) P B D} \nn \\
              && -   {\cal F}^{(+)}_{M A B} {\cal F}^{(+) M A}{}_{D} {\cal F}^{(+)}_{P E}{}^B {\cal F}^{(+) P E D} \ . \label{Rp}
\eea
The three contributions to the Lagrangian are generalized diffeomorphism scalars (modulo the strong constraint (\ref{StrongConstraint})), and the full Lagrangian  is $H$-invariant to first order in $\alpha'$
\be
\delta \left({\cal R}+ a {\cal R}^{(-)} + b {\cal R}^{(+)} \right) = \widehat {\cal L}_\xi \left({\cal R}+ a {\cal R}^{(-)} + b {\cal R}^{(+)} \right)  \ .
\ee
In fact, one can show that the anomalous Lorentz behaviour $\widetilde \delta_\Lambda {\cal R}$ is exactly cancelled by $\delta_\Lambda \left(a \, {\cal R}^{(-)} + b \, {\cal R}^{(+)}\right)$. We have verified this explicitly using \cite{Peeters:2007wn}. Notice also that $\widetilde \delta_\Lambda \left(a \, {\cal R}^{(-)} + b \, {\cal R}^{(+)}\right)$ is of higher order, so must not be considered in this computation. We then conclude that the action (\ref{ActionAlphaPrime}) is invariant under the $H$ and $\widehat {\cal L}$ symmetries.

Regarding $G$-symmetry, recall that in DFT the $O(d,d)$ transformations
\be
h_M{}^P \eta_{P Q} h_N{}^Q = \eta_{M N} \ ,
\ee
act as follows
\be
E_{M}{}^A \to h_M{}^P E_{P}{}^A  \ , \ \ \ \ \partial_M \to h_M{}^P \partial_P
\ .
\ee
Then, the action is manifestly $O(d,d)$ invariant since
all indices are contracted with the duality invariant metric.
Note however that if one chooses an $H$-gauge-fixed parameterization of the generalized frame (as we will do in the next section), a compensating $H$-transformation is required to restore the gauge. This
is no problem, as we have seen, because $H$ is a symmetry of the theory.

Let us finally mention that under the $Z_2$-parity transformation we find
\be
Z_2 \left({\cal R}^{(\pm)}\right) = {\cal R}^{(\mp)} \ ,
\ee
so again we see that the corrections are even under $Z_2$-parity
for $a = b$, odd for $a = -b$, and the parity is broken for any other choice.

 \subsection{Parameterization and field redefinitions}

Until now we have been general, and have  assumed neither a parameterization of the generalized fields nor any solution to the strong constraint (\ref{StrongConstraint}). Here we give the parameterizations required to make contact with the  deformed Bergshoeff-de Roo form of the action (\ref{BergshoeffdeRoo}).

The $G$-invariant metric is chosen to be as usual
 \be
\eta_{M N} = \left(\begin{matrix} 0 & \delta^\mu_\nu \\ \delta^\nu_\mu & 0 \end{matrix}\right) \ ,
\ee
and we choose the standard solution to the strong constraint for which
\be
\partial_M = (\tilde \partial^{\mu} , \partial_\mu) = \left(0 , \partial_\mu\right) \ . \label{SolSC}
\ee
 The flat metrics are parameterized as
\be
{\cal H}_{A B} =  \left(\begin{matrix}g^{a b} & 0 \\ 0 & g_{a b}\end{matrix}\right) \ , \ \ \ \ \ \eta_{A B} = \left(\begin{matrix}g^{a b} & 0 \\ 0 & - g_{a b} \end{matrix}\right) \ ,
\ee
and they are left invariant by $H$-transformations parameterized by
\be
\Lambda_A{}^B = \left( \begin{matrix} \Lambda^{(+)}{}^a{}_b & 0 \\ 0 & \Lambda^{(-)}{}_a{}^b\end{matrix}\right) \ .
\ee
Here, $\Lambda^{(+)}$ and $\Lambda^{(-)}$ are the Lorentz parameters that generate the $O(1,d-1)$  and $O(d-1,1)$-transformations that leave $\bar P_{A B}$ and $P_{A B}$ invariant respectively, and as such satisfy
\be
\Lambda^{(\pm)}_{a b} = g_{a c} \Lambda^{(\pm)}{}^c{}_b = - \Lambda^{(\pm)}_{b a} \ .
\ee

The generalized frame is parameterized by two beins $\bar e^{(\pm)}_\mu{}^a$ and a two-form $\bar B_{\mu \nu}$
\be
E_M{}^A = \frac{1}{\sqrt{2}} \left(\begin{matrix} \bar e^{(+)}_a{}^\mu & - g^{a b} \bar e^{(-)}_b{}^\mu \\ \bar e^{(+)}_\mu{}^b g_{b a} - \bar e^{(+)}_a{}^\rho \bar B_{\rho \mu} & \bar e^{(-)}_\mu{}^a + g^{a b} \bar e^{(-)}_b{}^\rho \bar B_{\rho \mu}\end{matrix}\right) \ .
\ee
The two beins satisfy
\be
\bar e^{(\pm)}_a{}^\mu \bar e^{(\pm)}_\mu{}^b = \delta^b_a \ , \ \ \ \ \ \bar e^{(\pm)}_\mu{}^a \bar e^{(\pm)}_a{}^\nu = \delta^\nu_\mu  \ , \ \ \ \ \ \bar e^{(\pm)}_{a}{}^\mu = \bar g^{\mu \nu} \bar e^{(\pm)}_\nu{}^b g_{b a} \ ,
\ee
and are constrained to reproduce the same symmetric metric $\bar g_{\mu \nu}$
\be
\bar g_{\mu \nu} = \bar e^{(\pm)}_{\mu}{}^a g_{a b} \bar e^{(\pm)}_{\nu}{}^b \ , \ \ \ \ \ \bar g^{\mu \nu} = \bar e^{(\pm)}_{a}{}^\mu g^{a b} \bar e^{(\pm)}_{b}{}^\nu \ .
\ee
They can be taken to be equal through a gauge fixing condition
\be
\bar e^{(+)}_\mu{}^b \, \Lambda^{(+)}_b{}^a = \bar e^{(-)}_\mu{}^b \, \Lambda^{(-)}_b{}^a = \bar e_\mu{}^b \Lambda_b{}^a\ ,
\ee
that breaks the $H$-group to the physical Lorentz group parameterized by $\Lambda_a{}^b$. The bars over the component fields indicate that they are first-order corrected, so for example
\be
\bar e_\mu{}^a = e_\mu{}^a + \alpha' \Delta e_\mu{}^a \ ,
\ee
where the un-barred part is of zeroth order, and transforms covariantly under diffeomorphisms and Lorentz transformations. However, the first order redefinition $\Delta e_\mu{}^a$ can induce a non-covariant behavior.

The matrices that generate the $Z_2$-parity transformations adopt the following parameterization
\be
Z_A{}^B = \left(\begin{matrix} 0 & g^{a b} \\ g_{a b} & 0 \end{matrix}\right) \ , \ \ \ \ \ Z_M{}^{N} = \left(\begin{matrix} - \delta^\mu_\nu & 0 \\ 0 & \delta_\mu^\nu \end{matrix}\right) \ ,
\ee
and at the level of components they exchange $Z_2 (\bar e^{(\pm)}_\mu{}^a ) = \bar e^{(\mp)}_\mu{}^a$. So, after the gauge fixing, they leave the bein (and thus the metric $\bar g_{\mu \nu}$) invariant, but they exchange the sign of the two-form $Z_2 (\bar B_{\mu \nu}) = - \bar B_{\mu \nu}$, as expected.

The generalized dilaton has the usual expression, which can be written either in terms of barred or un-barred fields
\be
e^{- 2 d} = \sqrt{- \bar g} e^{- 2 \bar \phi} = \sqrt{-  g} e^{- 2 \phi} \ . \label{gendilaton}
\ee
This is due to the fact that its gauge transformation (\ref{vard}) receives no first order correction. The equation (\ref{gendilaton}) defines the corrected dilaton $\bar \phi = \phi + \frac 1 4 \log \frac{\bar g} {g}$.
The generalized metric is parameterized as usual, but with respect to the barred fields
\be
{\cal H}_{M N} = \left( \begin{matrix} \bar g^{\mu \nu} & - \bar g^{\mu \rho} \bar B_{\rho \nu} \\ \bar B_{\mu \rho} \bar g^{\rho \nu} & \bar g_{\mu \nu} - \bar B_{\mu \rho} \bar g^{\rho \sigma} \bar B_{\sigma \nu}\end{matrix} \right) \ . \label{Paramgenmet}
\ee

The generalized fluxes appear in the action in terms that are purely of ${\cal O}(\alpha')$. This means that we only need their lowest order expressions in terms of the usual bein and two-form, i.e. we can drop the bars from these fields. Their four components are given by
\bea
{\cal F}_{a b c} &=& \sqrt{2} \omega^{(-)}_{\mu[bc} e_{a]}{}^\mu + \frac 1 {\sqrt{2}} \omega^{(+)}_{\mu[bc} e_{a]}{}^\mu \, ,\\
{\cal F}_{a b}{}^c &=& \frac 1 {\sqrt{2}} \omega^{(-)}_{\mu a b} g^{\mu \nu} e_\nu{}^c \, ,\\
{\cal F}_{a}{}^{b c} &=& - \frac 1 {\sqrt{2}} \omega^{(+) b c}_{\mu} e_a{}^\mu \, ,\\
{\cal F}^{a b c} &=& - \frac 1 {\sqrt{2}} \omega^{(-)[bc}_\mu e_\nu{}^{a]} g^{\mu \nu} - \sqrt{2} \omega^{(+)[bc}_\mu e_\nu{}^{a]} g^{\mu \nu} \ .
\eea
The projected fluxes can be written in components as well. We find that some projections vanish
\be
{\cal F}^{(-)}_{M a}{}^b = 0 \ , \ \ \ \ {\cal F}^{(-)}_{M}{}^{a b} = 0 \ , \ \ \ \ {\cal F}^{(+)}_{M a}{}^b = 0 \ , \ \ \ \ {\cal F}^{(+)}_{M a b} = 0  \ , \label{vanishingprojF}
\ee
leaving only the following non-vanishing components
\bea
{\cal F}^{(-)}_{M a b} &=& \frac 1 {\sqrt{2}}\left(\begin{matrix} e_c{}^\mu {\cal F}_{a b}{}^c \\ B_{\mu \nu} e_c{}^\nu {\cal F}_{a b}{}^c + e_\mu{}^d g_{d c} {\cal F}_{a b}{}^c\end{matrix}\right) = \frac 1 2 \left(\begin{matrix} g^{\mu \nu} \omega^{(-)}_{\nu a b} \\ B_{\mu \nu} g^{\nu \rho} \omega^{(-)}_{\rho a b}  + \omega^{(-)}_{\mu a b}\end{matrix}\right) \, ,\label{projfluxes}\\
{\cal F}^{(+) b c}_{M} &=& \frac 1 {\sqrt{2}}\left(\begin{matrix} - e_d{}^\mu g^{d a} {\cal F}_{a}{}^{b c} \\ - B_{\mu \nu} e_d{}^\nu g^{d a} {\cal F}_{a}{}^{b c} + e_\mu{}^a {\cal F}_{a}{}^{b c}\end{matrix}\right) =  \frac 1 2 \left(\begin{matrix} g^{\mu \nu} \omega^{(+)b c}_{\nu} \\ B_{\mu \nu} g^{\nu \rho} \omega^{(+) b c}_{\rho}  - \omega^{(+) b c}_{\mu}\end{matrix}\right) \ . \nn
\eea

Now that we have parameterized all the generalized fields, we study the behavior of the components under generalized transformations. The action (\ref{ActionAlphaPrime}) depends only on the generalized metric and the projected fluxes, so we will only focus on the transformations of these objects. Regarding the projected fluxes, as we explained only their lowest order terms are relevant to ${\cal O}(\alpha')$ and it can be easily verified that the transformations (\ref{varprojectedF}) reproduce the expected transformations for their components (see for example \cite{Geissbuhler:2013uka}). The transformation of the generalized metric instead requires a special treatment, as its first order correction plays a fundamental role in this construction. When the parameterization (\ref{Paramgenmet}) is subjected to the transformation (\ref{vargenmet}) restricted to the choice (\ref{SolSC}), the components of the generalized metric transform as
\bea
\delta \bar g_{\mu \nu} &=& L_\xi \bar g_{\mu \nu} - \frac a 2 \omega^{(-)}_{(\mu a}{}^b \partial_{\nu)} \Lambda_b{}^a  - \frac b 2 \omega^{(+)}_{(\mu a}{}^b \partial_{\nu)} \Lambda_b{}^a \, ,\label{compGS1}\\
\delta \bar B_{\mu \nu} &=& L_\xi \bar B_{\mu \nu} + 2 \partial_{[\mu}\xi_{\nu]} + \frac a 2 \omega^{(-)}_{[\mu a}{}^b \partial_{\nu]} \Lambda_b{}^a  - \frac b 2 \omega^{(+)}_{[\mu a}{}^b \partial_{\nu]} \Lambda_b{}^a \ . \label{compGS2}
\eea
We then see that the generalized Green-Schwarz transformation (\ref{genGSgenmetric}) affects not only the two-form, but also the metric. They both receive a
non-covariant Lorentz transformation. In order to relate $\bar g_{\mu \nu}$ to the usual Lorentz-singlet metric $g_{\mu \nu}$ that transforms covariantly, a first-order in $\alpha'$ Lorentz non-covariant field redefinition is required
\be
\bar g_{\mu \nu} = g_{\mu \nu} - \frac a 4 \omega^{(-)}_{\mu a}{}^b \omega^{(-)}_{\nu b}{}^a - \frac b 4 \omega^{(+)}_{\mu a}{}^b \omega^{(+)}_{\nu b}{}^a  \ . \label{fieldredefg}
\ee
For generic values of the parameters $(a,b)$ such a redefinition of the two-form is not possible. We will comment on this point at the end of this section, and by now let us simply mention that in the component action that we write down below, $\bar B = B^{BR}$.

Introducing the non-vanishing components of the
projected fluxes  (\ref{projfluxes}) and the generalized metric  (\ref{Paramgenmet}) into (\ref{ActionAlphaPrime}), and performing the field redefinition (\ref{fieldredefg}), we can finally write the Lagrangian in components (we have benefited from \cite{Peeters:2007wn} in this computation)
\bea
{\cal R} + a {\cal R}^{(-)} + b {\cal R}^{(+)} &=& R - 4 \nabla_\mu \phi \nabla^\mu \phi + 4 \nabla_\mu \nabla^\mu \phi - \frac 1 {12} \widetilde H^{\mu \nu \rho} \widetilde H_{\mu \nu \rho} \nn\\
&& + \frac a 8 R^{(-)}_{\mu \nu a}{}^b R^{(-)\mu \nu}{}_b{}^a + \frac b 8 R^{(+)}_{\mu \nu a}{}^b R^{(+)\mu \nu}{}_b{}^a \ , \label{componentaction}
\eea
where
\be
\widetilde H_{\mu \nu \rho} = H_{\mu \nu \rho} - \frac 3 2 a \Omega^{(-)}_{\mu \nu \rho} + \frac 3 2 b \Omega^{(+)}_{\mu \nu \rho} \ .
\ee
Written in this way, the invariance under the following transformations to ${\cal O}(\alpha')$ is manifest
\bea
\delta \phi &=& L_\xi \phi \, ,\\
\delta  g_{\mu \nu} &=& L_\xi g_{\mu \nu}  \, ,\\
\delta B_{\mu \nu} &=& L_\xi  B_{\mu \nu} + 2 \partial_{[\mu}\xi_{\nu]} + \frac a 2 \omega^{(-)}_{[\mu a}{}^b \partial_{\nu]} \Lambda_b{}^a  - \frac b 2 \omega^{(+)}_{[\mu a}{}^b \partial_{\nu]} \Lambda_b{}^a \ .
\eea
The action (\ref{componentaction}) exactly coincides with the
two-parameter deformations of the Bergshoeff-de Roo form of the action (\ref{BergshoeffdeRoo}).
We have then re-formulated such deformations in an $O(d,d)$-invariant way (\ref{ActionAlphaPrime}).

Let us conclude this section with some remarks. We have seen that the generalized metric is $\alpha'$-corrected, but it is still symmetric and $O(d,d)$-valued, and as such can be parameterized as in (\ref{Paramgenmet}). The barred fields $\bar g_{\mu \nu}$ and $\bar B_{\mu \nu}$ are duality covariant, but the generalized Green-Schwarz transformation induces the Lorentz non-covariant transformations (\ref{compGS1}), (\ref{compGS2}) of these duality covariant components. In the case of the metric, we have shown how a Lorentz non-covariant field redefinition (\ref{fieldredefg}) related the duality covariant metric $\bar g_{\mu \nu}$ with the standard Lorentz-singlet covariant metric $g_{\mu \nu}$. For generic values of the parameters such a redefinition is not possible for $\bar B_{\mu \nu}$. This was expected since a given choice of parameters reproduces the heterotic string, in which the two-form necessarily acquires the anomalous Lorentz transformation required for anomaly cancellations in the Green-Schwarz mechanism. However, the
two-form in the  closed bosonic string must be a Lorentz-singlet, so when $a = b$ we should be able to remove the Lorentz non-covariant behavior of the two-form through some non-covariant field redefinition. When $a = b$, the redefinition of the metric (\ref{fieldredefg}) becomes
\be
\bar g_{\mu \nu} = g_{\mu \nu} - \frac a 2 \omega_{\mu a}{}^b \omega_{\nu b}{}^a - \frac a 8 H_{\mu a}{}^b H_{\nu b}{}^a \ . \label{redefbosg}
\ee
Regarding the two-form, when $a = b$ its anomalous transformation (\ref{compGS2})
\bea
\delta_\Lambda \bar B_{\mu \nu} &=& - \frac a 2 H_{[\mu a}{}^b \partial_{\nu]} \Lambda_b{}^a \ ,
\eea
can
be removed in this case through a $Z_2$-parity-preserving Lorentz non-covariant field redefinition
\be
\bar B_{\mu \nu} = B_{\mu \nu} - \frac a 2 H_{[\mu a}{}^b \omega_{\nu] b}{}^a \ .\label{redefbosB}
\ee
The redefinitions (\ref{redefbosg}) and (\ref{redefbosB}) then take the form of background-independent Lorentz non-covariant versions of Meissner's field redefinitions \cite{Meissner:1996sa}. Then, while $g_{\mu \nu}$ and $B_{\mu \nu}$ are diffeomorphism and Lorentz covariant, $\bar g_{\mu \nu}$ and $\bar B_{\mu \nu}$ are Lorentz non-covariant but T-duality covariant.

Regarding the heterotic case $(a,b)=(- \alpha', 0)$, our results predict the field redefinitions of \cite{Bergshoeff:1995cg} that relate the Lorentz-covariant metric with the T-duality covariant one. In addition, we obtain the anomalous Lorentz transformation of the metric as given in \cite{ht}, plus the usual Green-Schwarz transformation of the two-form in terms of $\omega^{(-)}$. Another interesting example in which the non-covariant Lorentz transformation of the two-form cannot be removed through a field redefinition is the case $(a,b)=(-\alpha', \alpha')$. This theory contains no Riemann squared terms, and the essential first order contributions are given by Chern-Simons corrections to the
curvature of the two-form. Then,
being the corrections
odd under $Z_2$-parity,
this case is similar to the one introduced in \cite{Hohm:2013jaa}, with the difference that the non-covariance in this case is due to Lorentz and in \cite{Hohm:2013jaa} it is due to diffeomorphisms.

\section{Outlook and concluding remarks} \label{Conclu}

We have shown that the four-derivative terms in the string effective actions admit a universal description in terms of a two-parameter family of theories. The two parameters ($a$ and $b$) interpolate between corrections that are even ($a = b$) and odd ($a=-b$) with respect to a parity transformation that exchanges the sign of the two-form. We have given two expressions for
the two-parameter deformed theory, which are related by field redefinitions.
One of them facilitates comparison with the closed bosonic and heterotic string effective actions as presented by R. Metsaev and A. Tseytlin in \cite{Metsaev:1987zx}, and the other one admits a direct comparison with the heterotic string
effective action as formulated by E. Bergshoeff and M. de Roo in \cite{Bergshoeff:1988nn}.
The action depends on the frame, two-form and dilaton fields, and we have neglected the contributions from the heterotic gauge fields for simplicity.

We have then reformulated the two-parameter action in the $O(d,d)$ invariant language of DFT. The first novel contribution is a first-order correction to the gauge transformations of the generalized fields that takes the form of a generalized Green-Schwarz transformation (see for example (\ref{genGSgenmetric})), that generically cannot be removed through a duality covariant generalized field redefinition. This anomalous Lorentz transformation implies that its field components also transform non-covariantly, as explicitly shown in
(\ref{compGS1})-(\ref{compGS2}). While this non-covariant behavior can be removed from the metric through a Lorentz non-covariant first-order field redefinition (\ref{fieldredefg}), this is not possible in general for the two-form. For example, when the parameters are chosen to reproduce the heterotic string effective action, the two-form receives the anomalous Lorentz transformation required for anomaly cancelation in the Green-Schwarz mechanism (which cannot be removed through field redefinitions). Instead, in the even parity case there is a Lorentz non-covariant redefinition of the two-form that renders it covariant, as expected for the closed
bosonic string.

The generalized Green-Schwarz transformation is also very powerful in that it gives rise to a duality covariant gauge principle that demands and  determines the first-order $\alpha'$-corrections in the action. The lowest order DFT action (\ref{DFTaction}) is invariant under generalized diffeomorphisms, but not under these novel higher-derivative Lorentz transformations. As a consequence, the four-derivative terms (\ref{Rm}) and (\ref{Rp}) must be added to the action in order
to cancel the anomalous transformation. When the strong constraint is solved in the (super)gravity frame and the generalized fields are parameterized accordingly, the resulting four-derivative action (\ref{componentaction}) receives contributions not only from the explicit four-derivative terms (\ref{Rm}) and (\ref{Rp}), but also from the two-derivative terms (\ref{DFTaction}) through the
first-order in $\alpha'$ redefinitions of the fields.
When
the component fields parameterizing the generalized fields are specified,
the final form of the
action exactly
coincides
with the two-parameter Bergshoeff-de Roo action discussed in section
\ref{SEC:BR}.

Similar results where obtained by O. Hohm and B. Zwiebach in
\cite{Hohm:2014xsa}. They constructed a
two-parameter $O(d,d)$ invariant theory up to cubic order in perturbations
of the fields, in which the parameters $\gamma^{(\pm)}$ interpolate between
even (DFT$^+$) and odd
(DFT$^-$) $Z_2$-parity
corrections. Their formulation is metric-like, and then all the fields are Lorentz invariant. The generalized gauge transformations do receive ${\cal O}(\alpha')$ corrections, which are generated by the generalized infinitesimal diffeomorphism parameter $\xi^M$. The duality covariant fields that appear as components of the $O(d,d)$ multiplets then transform non-covariantly under diffeomorphisms, rather than Lorentz transformations. Although this is different from the approach
we have followed here, it is possible that both formulations can be related through local (generalized) field redefinitions like the ones explored in \cite{Hohm:2014eba}. The similarity between both approaches is evident to the point that it is natural to identify the parameters $\gamma^{(\pm)} = -\frac{ a \pm b} 2$.

Our work is essentially an $O(d,d)$ invariant re-writing of the first order
$\alpha'$-corrections in the string effective actions. At the moment it is unclear if this formulation admits an extension to higher orders.  An important application of this line of research would be to
find a duality covariant gauge principle that  requires and fixes the higher-derivative
terms in the $\alpha'$-expansion, as it could provide a tool to compute corrections that are otherwise
difficult to calculate through other methods. A less ambitious programme that
could give hints on how to proceed in this direction is to rewrite the
already known higher derivative ($\alpha'^{n}, n=2,3,4$) corrections to the string effective actions in an $O(d,d)$ invariant way.

Other possible directions for future work suggest themselves. It is possible that our formulation admits a description in terms of an extended-tangent space formulation like the one considered in \cite{Bedoya:2014pma}-\cite{Lee:2015kba}, in which the tangent space should be further enhanced so as to include two spin connections with opposite torsion with duality group $O(d+n,d+n)$. Understanding the role of supersymmetry would also be of interest, since one should expect obstructions when attempting to supersymmetrize this theory for a
choice of parameters leaving only even $Z_2$-parity corrections.  Generalized Scherk-Schwarz reductions like those considered in \cite{Aldazabal:2011nj} would also be interesting to examine in order to find higher-derivative corrections in gauged supergravities and to clarify the relation between $\alpha'$-corrections and non-geometry (see for example \cite{Andriot:2012an} and references therein). Due to the field redefinitions involved in this construction, we expect the duality covariant scalars of the reduced theory to be related to the diffeomorphism and Lorentz covariant scalars through  ${\cal O}(\alpha')$ redefinitions that are quadratic in gaugings. A pure generalized flux formulation of the theory \cite{Geissbuhler:2013uka} could be useful in understanding these issues. Finally, the generalized Green-Schwarz transformation might be relevant in the analysis of large gauge transformations in DFT \cite{Hohm:2012gk}.

~

\noindent {\bf \underline{Acknowledgments:}} We are indebted with O. Hohm and B. Zwiebach for many enlightening discussions and comments on the draft, and we also warmly thank O. Bedoya and
J.J. Fern\'andez Melgarejo. We wish to thank O. Hohm for pointing out a mistake in eq. (3.35) of v1, which has now been corrected.
D. M. thanks the Center for Theoretical Physics at MIT  for kind hospitality
during the early stages of this work. Support by the Fulbright Commission,
 A.S.ICTP,
CONICET, UBA and ANPCyT is also gratefully acknowledged.

~

\begin{appendix}
\section{Conventions and definitions}\label{SEC:Conventions}
In this Appendix we introduce the notation used throughout the paper.

Space-time and tangent space Lorentz indices are denoted $\mu, \nu, \dots$
and $a,b,\dots$, respectively. The Lie derivative of a tensor is given by
\be
L_{\xi} V_{\mu}{}^{\nu}= \xi^{\rho} \partial_{\rho}V_{\mu}{}^{\nu} + \partial_{\mu} \xi^{\rho} V_{\rho}{}^{\nu} - \partial_{\rho} \xi^{\nu} V_{\mu}{}^{\rho} \ .
\ee
The Christoffel connection is defined in terms of the metric as
\be
\Gamma_{\mu \nu}^{\rho} = \frac 1 2 g^{\rho \sigma} \left( \partial_{\mu} g_{\nu \sigma} + \partial_{\nu} g_{\mu \sigma} - \partial_{\sigma} g_{\mu \nu}\right) \ , \ \ \ \ \ \ \Gamma_{[\mu \nu]}^{\rho} = 0 \ ,
\ee
and transforms anomalously under infinitesimal diffeomorphisms (whenever the Lie derivative acts on a
non-tensorial object, we use the convention that it acts as if it
were covariant)
\be
\delta_{\xi} \Gamma_{\mu \nu}^\rho = L_{\xi}  \Gamma_{\mu \nu}^\rho + \partial_{\mu} \partial_{\nu} \xi^{\rho} \ ,
\ee
so it allows to define a covariant derivative, given by
\be
\nabla_{\rho} V_{\mu}{}^{\nu} = \partial_{\rho} V_{\mu}{}^{\nu} - \Gamma_{\rho \mu}^{\sigma} V_{\sigma}{}^{\nu} + \Gamma_{\rho \sigma}^{\nu} V_{\mu}{}^{\sigma} \ .
\ee
The Riemann tensor can be expressed as
\be
R^{\rho}{}_{\sigma \mu \nu} = \partial_{\mu} \Gamma_{\nu \sigma}^{\rho} - \partial_{\nu} \Gamma_{\mu \sigma}^\rho + \Gamma_{\mu \delta}^\rho \Gamma_{\nu \sigma}^\delta - \Gamma_{\nu \delta}^\rho \Gamma_{\mu \sigma}^\delta \ . \label{curvedRiemann}
\ee
Its  symmetries and Bianchi identities are
\be
R_{\rho \sigma \mu \nu} = g_{\rho \delta} R^{\delta}{}_{\sigma \mu \nu} = R_{([\rho \sigma][\mu \nu])}
\ , \ \ \ \
R^{\rho}{}_{[\sigma \mu \nu]} = 0 \ , \ \ \ \ \nabla_{[\mu} R_{\nu \lambda]}{}^\rho{}_\sigma = 0 \ .
\ee
Traces of the Riemann tensor give the Ricci tensor and scalar, respectively
\be
R_{\mu \nu} = R^{\rho}{}_{\mu \rho \nu} \ , \ \ \ \ \ \ R = g^{\mu \nu} R_{\mu \nu} \ .
\ee
The (inverse) metric can be written in terms of a (inverse) frame field
\be
g_{\mu \nu} = e_{\mu}{}^a g_{a b} e_{\nu}{}^b \ , \ \ \ \ \ g^{\mu \nu} = e_{a}{}^\mu g^{a b} e_{b}{}^\nu \ ,
\ee
where $g_{a b}$ is the Minkowski metric, and they satisfy the following identities
\be
e_a{}^\mu e_\mu{}^b = \delta^b_a \ , \ \ \ \ \ e_\mu{}^a e_a{}^\nu = \delta^\nu_\mu
\ , \ \ \ \ \ e_{a}{}^\mu = g^{\mu \nu} e_\nu{}^b g_{b a} \ .
\ee
Under Lorentz and infinitesimal diffeomorphism transformations, the frame
field changes as follows
\be
\delta e_{\mu}{}^a = L_\xi e_{\mu}{}^a + e_{\mu}{}^b \Lambda_b{}^a \ , \ \ \ \ \
\delta e_{a}{}^\mu = L_\xi e_{a}{}^\mu - \Lambda_a{}^b e_{b}{}^\mu
 \ , \ \ \ \ \
\Lambda_{a b} = \Lambda_a{}^c g_{c b} = - \Lambda_{b a} \ .
\ee
We also consider a spin connection defined in terms of the frame field
\be
\omega_{\mu a}{}^b = \partial_{\mu} e_{\nu}{}^b e_a{}^\nu - \Gamma_{\mu \nu}^\rho e_\rho{}^b e_a{}^\nu \ , \label{spinconnection}
\ee
that transforms as
\be
\delta \omega_{\mu a}{}^b = L_\xi \omega_{\mu a}{}^b + \partial_\mu \Lambda_{a}{}^b + \omega_{\mu a}{}^c \Lambda_c{}^b - \Lambda_a{}^c \omega_{\mu c}{}^b \ .
\ee
Given a Lorentz tensor
\be
\delta_\Lambda T_{a}{}^b = T_a{}^c \Lambda_c{}^b - \Lambda_a{}^c T_c{}^b \ ,
\ee
we define the Lorentz covariant derivative
\be
   {\cal D}_\mu T_a{}^b = \partial_\mu T_a{}^b + \omega_{\mu a}{}^c T_c{}^b -
   \omega_{\mu c}{}^b T_a{}^c \ .
\ee
The Riemann tensor can also be written as an adjoint Lorentz-valued two-form,
expressed in terms of the spin connection as
\be
R_{\mu \nu a}{}^b = \partial_{\mu} \omega_{\nu a}{}^b - \partial_{\nu}
\omega_{\mu a}{}^b + \omega_{\mu a}{}^c \omega_{\nu c}{}^b  -   \omega_{\nu a}{}^c
\omega_{\mu c}{}^b \ . \label{twoformRiemann}
\ee
This form of the Riemann tensor transforms as
\be
\delta R_{\mu \nu a}{}^b = L_\xi R_{\mu \nu a}{}^b + R_{\mu \nu a}{}^c \Lambda_c{}^b -
\Lambda_a{}^c R_{\mu \nu c}{}^b \ ,
\ee
and is related to the  Riemann tensor (\ref{curvedRiemann}) through a
frame rotation
\be
R_{\mu \nu a}{}^b e_{b}{}^\rho e_\sigma{}^a = - R^{\rho}{}_{\sigma \mu \nu} \ .
\ee
The Lorentz and diffeomorphism covariant derivatives are related as follows
\be
   {\cal D}_\mu T_a{}^b = \nabla_\mu T_\rho{}^\sigma e_a{}^\rho e_\sigma{}^b  \ \ \ \ \
   \ \ {\rm for} \ \ \ \ \ \ \ T_a{}^b = T_\rho{}^\sigma e_a{}^\rho e_\sigma{}^b \ .
\ee
The Chern-Simons three-form is defined as
\be
\Omega_{\mu \nu \rho} = \omega_{[\mu a}{}^b \partial_{\nu} \omega_{\rho] b}{}^a + \frac 2 3
\omega_{[\mu a}{}^b \omega_{\nu b}{}^c \omega_{\rho] c}{}^a \ ,
\ee
and it transforms under infinitesimal diffeomorphisms and Lorentz transformations as
\be
\delta \Omega_{\mu \nu \rho} = L_\xi \Omega_{\mu \nu \rho} -
\partial_{[\mu} \left(\partial_{\nu} \Lambda_a{}^b \omega_{\rho] b}{}^a\right) \ .
\ee
We also define the spin connections with torsion
\be
\omega^{(\pm)}_{\mu a}{}^b = \omega_{\mu a}{}^b \pm \frac 1 2 H_{\mu a}{}^b \ , \ \ \ \ \ \ H_{\mu a}{}^b = H_{\mu \nu \rho} e_a{}^\nu g^{\rho \sigma} e_\sigma{}^b \ ,
\ee
where the torsion is given by the three form curvature of the Kalb-Ramond two-form
\be
H_{\mu \nu \rho} = 3 \partial_{[\mu} B_{\nu \rho]} = \partial_{\mu} B_{\nu \rho} + \partial_{\nu} B_{\rho \mu} + \partial_{\rho} B_{\mu \nu} \ ,
\ee
with Bianchi identity
\be
\nabla_{[\mu} H_{\nu \rho \sigma]} = 0 \ .
\ee
Note that we do not include any $\alpha'$-correction in the torsion, as we are only interested in first-order corrections in this paper. We also define powers of the three-form with the following contractions
\be
H^4 = H^{\mu \nu \rho} H_{\mu \sigma}{}^\lambda H_{\nu \lambda}{}^\delta H_{\rho \delta}{}^\sigma \ , \ \ \ \ H^2_{\mu \nu} = H_{\mu}{}^{\rho \sigma} H_{\nu \rho \sigma} \ , \ \ \ \ H^{2} = H_{\mu \nu \rho} H^{\mu \nu \rho} \ .
\ee
When the two-form Riemann tensor is supra-labeled with a sign, we use the convention that it is defined as in (\ref{twoformRiemann}) but
in terms of the spin connection with torsion
\be
R^{(\pm)}_{\mu \nu a}{}^b = \partial_{\mu} \omega^{(\pm)}_{\nu a}{}^b - \partial_{\nu} \omega^{(\pm)}_{\mu a}{}^b + \omega^{(\pm)}_{\mu a}{}^c \omega^{(\pm)}_{\nu c}{}^b  -   \omega^{(\pm)}_{\nu a}{}^c \omega^{(\pm)}_{\mu c}{}^b \ .
\ee
The supra-labeled
with a sign torsionful Chern-Simons three-form is accordingly
\be
\Omega^{(\pm)}_{\mu \nu \rho} = \omega^{(\pm)}_{[\mu a}{}^b
  \partial_{\nu} \omega^{(\pm)}_{\rho] b}{}^a + \frac 2 3 \omega^{(\pm)}_{[\mu a}{}^b \omega^{(\pm)}_{\nu b}{}^c \omega^{(\pm)}_{\rho] c}{}^a \ .
\ee
The transformations of the torsionful spin connection, Riemann tensor and Chern-Simons three-form are as follows
\bea
\delta \omega^{(\pm)}_{\mu a}{}^b &=& L_\xi \omega^{(\pm)}_{\mu a}{}^b + \partial_\mu \Lambda_{a}{}^b + \omega^{(\pm)}_{\mu a}{}^c \Lambda_c{}^b - \Lambda_a{}^c \omega^{(\pm)}_{\mu c}{}^b \, ,\\
\delta R^{(\pm)}_{\mu \nu a}{}^b &=& L_\xi R^{(\pm)}_{\mu \nu a}{}^b + R^{(\pm)}_{\mu \nu a}{}^c \Lambda_c{}^b - \Lambda_a{}^c R^{(\pm)}_{\mu \nu c}{}^b\, ,\\
\delta \Omega^{(\pm)}_{\mu \nu \rho} &=& L_\xi \Omega^{(\pm)}_{\mu \nu \rho} - \partial_{[\mu} \left(\partial_{\nu} \Lambda_a{}^b \omega^{(\pm)}_{\rho] b}{}^a\right) \ .
\eea

\section{From Bergshoeff-de Roo to Metsaev-Tseytlin} \label{SEC:Comparison}
It is very easy to show that the two-parameter generalization of the Bergshoeff-de Roo action (\ref{BergshoeffdeRoo}) is equivalent to the
two-parameter deformation of the Metsaev-Tseytlin action (\ref{ActionMT})-(\ref{MetsaevTseytlin}), up to field redefinitions and boundary terms.
For the heterotic case $(a,b) = (- \alpha',0)$, the equivalence
was proved in \cite{Chemissany:2007he}, and here we give the general proof for arbitrary values of the coefficients\footnote{Field redefinitions in the context of the heterotic string have also been discussed recently in \cite{delaOssa:2014msa}.}. The zeroth order actions are automatically identical (both given by (\ref{L0})), so we need to focus attention on the four-derivative corrections.

Using the decomposition of the Riemann tensor with torsion
\be
R^{(\pm)}_{\mu \nu a b} = R_{\mu \nu a b} \pm {\cal D}_{[\mu} H_{\nu] a b} - \frac 1 2 H_{[\mu a}{}^{c} H_{\nu] b c} \ ,
\ee
 the components of the torsionful Riemann squared terms are
\bea
\frac a 8 R^{(-)}_{\mu \nu a}{}^b R^{(-)\mu \nu}{}_b{}^a + \frac b 8 R^{(+)}_{\mu \nu a}{}^b R^{(+)\mu \nu}{}_b{}^a &=& - \frac 1 {8} (a + b) \left[  {\cal D}_{[\mu} H_{\nu] a b} {\cal D}^\mu H^{\nu a b} + R_{\mu \nu a b} R^{\mu \nu a b} \right. \nn \\
&&\ \ \ \ \ \ \ \   -  H_{\mu a}{}^c H_{\nu b c} R^{\mu \nu a b} + \frac 1 {8} H_{\mu a c} H^{\mu a}{}_d H_{\nu b}{}^c H^{\nu b d} \nn \\
&&\ \ \ \ \ \ \ \  \left. - \frac 1 {8}  H_{\mu a b} H^{\mu}{}_{c d} H_{\nu}{}^{a c} H^{\nu b d} \right] \label{DecompR2} \\
&& + \frac 1 4 (a - b) \left[ {\cal D}_{\mu}H_{\nu a b} R^{\mu \nu a b}  - \frac 1 2 {\cal D}_{\mu} H_{\nu a b} H^{\mu a}{}_c H^{\nu b c} \right] \ . \nn
\eea
On the other hand,  consider the first order in the decomposition of the squared three-form term
\bea
\left[- \frac 1 {12} \widetilde H^{\mu \nu \rho} \widetilde H_{\mu \nu \rho}\right]^{(1)} &=&  \frac 1 8 (a + b) H^{\mu \nu \rho} \left[   \partial_\mu \left(H_{\nu}{}^{a b} \omega_{\rho a b}\right) +   H_{\mu a b} R_{\nu \rho}{}^{a b} - \frac 1 {6} H_{\mu a}{}^b H_{\nu b}{}^c H_{\rho c}{}^a\right] \nn \\
&& + \frac 1 4 (a - b) H^{\mu \nu \rho} \left[ \Omega_{\mu \nu \rho} - \frac 1 {4} {\cal D}_{\mu} H_{\nu}{}^{a b} H_{\rho a b}\right] \ . \label{DecomptH2}
\eea
Now, using Bianchi identities one can show that the following terms vanish
\be
{\cal D}_{\mu}H_{\nu a b} R^{\mu \nu a b} = 0 \ , \ \ \ \ {\cal D}_{\mu} H_{\nu a b} H^{\mu a}{}_c H^{\nu b c} = 0 \ , \ \ \ \ H^{\mu \nu \rho} {\cal D}_{\mu} H_{\nu}{}^{a b} H_{\rho a b} = 0 \ , \label{VanishingId}
\ee
so (\ref{DecompR2}) only depends on $a + b$ and is then even under $Z_2$-parity,  and also prove the following useful identities
\be
H_{\mu \rho}{}^\lambda H_{\nu \sigma \lambda} R^{\mu \nu \rho \sigma} = \frac 1 2 H_{\mu \rho}{}^\lambda H_{\nu \sigma \lambda} R^{\mu \rho \nu \sigma} \ , \ \ \ \ \ \ \nabla_{[\mu} H_{\nu] \rho \sigma} \nabla^\mu H^{\nu \rho \sigma} = \frac 1 3 \nabla_{\mu} H_{\nu \rho \sigma} \nabla^\mu H^{\nu \rho \sigma}  \ . \label{Identities}
\ee
Adding (\ref{DecompR2}) and (\ref{DecomptH2}), canceling the terms in (\ref{VanishingId}) and rewriting some terms as in (\ref{Identities}), we find the
first order component of the Bergshoeff-de Roo Lagrangian
\bea
L^{(1)} &=& \frac 1 4 (a - b) H^{\mu \nu \rho}  \Omega_{\mu \nu \rho} + \frac 1 8 (a + b) H^{\mu \nu \rho}  \partial_\mu \left(H_{\nu}{}^{a b} \omega_{\rho a b}\right) \nn \\
&&  -\frac 1 8 (a + b) \left[ R_{\mu \nu \rho \sigma} R^{\mu \nu \rho \sigma} - \frac 3 2  H^{\mu \nu \rho}   H_{\mu \sigma \lambda} R_{\nu \rho}{}^{\sigma \lambda} + \frac 1 {24} H^{\mu \nu \rho} H_{\mu \sigma}{}^\lambda H_{\nu \lambda}{}^\delta H_{\rho \delta}{}^\sigma \nn \right.\\
&& \left. \ \ \ \ \ \ \ \ \ \ \ \ \ \ \ \ + \frac 1 3 \nabla_{\mu} H_{\nu \rho \sigma} \nabla^\mu H^{\nu \rho \sigma} + \frac 1 {8} H_{\mu \rho \delta} H^{\mu \rho}{}_\lambda H_{\nu \sigma}{}^\delta H^{\nu \sigma \lambda} \right] \ .\label{L1partial}
\eea
The first term in (\ref{L1partial}) is the Chern-Simons term present in the Metsaev-Tseytlin form of the action (\ref{MetsaevTseytlin}). The second term can be simply removed by a Lorentz non-covariant field redefinition of the two-form. The last block of terms with coefficient $a + b$ is even under $Z_2$-parity,
and exactly agrees with the results in \cite{godazgar},  where it was shown  to coincide modulo field redefinitions and boundary terms with the
Metsaev-Tseytlin form of the action \cite{Metsaev:1987zx}. In order to make contact with it, we note that
\bea
\left[L^{(0)}\left(g + \Delta g , B + \Delta B , \phi + \Delta \phi \right)\right]^{(1)} \!\! &=& \! \! e^{2 \phi}\nabla_\mu \left(e^{-2\phi} V^\mu \right) \label{L0to1}\\
&& \! \! + \frac {a + b} 8 H^{\mu \nu \rho}   H_{\mu \sigma \lambda} R_{\nu \rho}{}^{\sigma \lambda} - \frac {a + b} {32} H_{\mu \rho \delta} H^{\mu \rho}{}_\lambda H_{\nu \sigma}{}^\delta H^{\nu \sigma \lambda} \nn\\&& \!\! - \frac {a + b} {24} \nabla_{\mu}H_{\nu \rho \sigma} \nabla^{\mu}H^{\nu \rho \sigma} + \frac {a + b} 8 H^{\mu \nu \rho} \partial_{\mu}\left(H_\nu{}^{a b} \omega_{\rho a b}\right) \ , \nn
\eea
with
\bea
\Delta g_{\mu \nu} &=& - \frac 1 8 (a + b) H_{\mu}{}^{\rho \sigma} H_{\nu \rho \sigma} \, ,\nn\\
\Delta B_{\mu \nu} &=&  - \frac 1 4 (a + b) \left( \nabla^\rho H_{\rho \mu \nu} - 2 \nabla_\rho \phi H^\rho{}_{\mu \nu}\right) - \frac 1 4 (a + b) H_{[\mu}{}^{a b} \omega_{\nu] a b} \, ,\label{deltaBnoncovMTvsBR}\\
\Delta \phi &=&  - \frac 1 {32} (a + b) H_{\mu \nu \rho} H^{\mu \nu \rho} \ ,\nn
\eea
and
\be
V^\mu = - \frac 1 8 (a + b) H^{\mu \rho \sigma} \left( \nabla_\nu H^\nu{}_{\rho \sigma} - 2 \nabla_\nu \phi H^\nu{}_{\rho \sigma}\right) \ . \label{Vboundary}
\ee
That is, a shift in the zeroth order Lagrangian (\ref{L0}) due to the first order field redefinitions (\ref{deltaBnoncovMTvsBR})
(which coincide with those in \cite{godazgar} for the choice of
parameters $(a,b) = (- \alpha', -\alpha')$ reproducing the bosonic string), produces a covariant boundary term defined by (\ref{Vboundary}), plus the additional terms in the last two lines in (\ref{L0to1}). These terms take the first order Lagrangian (\ref{L1partial}) to the form
\bea
L^{(1)} &=& \frac 1 4 (a - b) H^{\mu \nu \rho}  \Omega_{\mu \nu \rho}  \\
&&  -\frac 1 8 (a + b) \left[ R_{\mu \nu \rho \sigma} R^{\mu \nu \rho \sigma} - \frac 1 2  H^{\mu \nu \rho}   H_{\mu \sigma \lambda} R_{\nu \rho}{}^{\sigma \lambda} \right. \nn \\ && \left. \ \ \ \ \ \ \ \ \ \ \ \ \ \ \ \ + \frac 1 {24} H^{\mu \nu \rho} H_{\mu \sigma}{}^\lambda H_{\nu \lambda}{}^\delta H_{\rho \delta}{}^\sigma - \frac 1 {8} H_{\mu \rho \delta} H^{\mu \rho}{}_\lambda H_{\nu \sigma}{}^\delta H^{\nu \sigma \lambda} \right] \ , \nn
\eea
which is exactly the first order correction in the two-parameter Metsaev-Tseytlin action (\ref{MetsaevTseytlin}).

Then, we have shown that the deformed Bergshoeff-de Roo action exactly coincides with the deformed Metsaev-Tseytlin action up to field redefinitions and boundary terms. We note that while the field redefinitions of the metric and dilaton are covariant, the redefinition of the two-form receives a Lorentz non-covariant contribution from the last term in $\Delta B_{\mu\nu}$ in (\ref{deltaBnoncovMTvsBR}).
\end{appendix}

\end{document}